# Disentangling amplitude and phase dynamics of a charge density wave in a photo-induced phase transition

Alfred Zong,[1,*] Anshul Kogar,[1,*] Ya-Qing Bie,[1] Timm Rohwer,[1,†] Changmin Lee,[1] Edoardo Baldini,[1] Emre Ergeçen,[1] Mehmet B. Yilmaz,[1] Byron Freelon,[1,‡] Edbert J. Sie,[1,§] Hengyun Zhou,[1,¶] Joshua Straquadine,[2,3,4] Philip Walmsley,[2,3,4] Pavel E. Dolgirev,[5] Alexander V. Rozhkov,[5,6] Ian R. Fisher,[2,3,4] Pablo Jarillo-Herrero,[1] Boris V. Fine,[5,7] and Nuh Gedik[1,**]

[1]*Massachusetts Institute of Technology, Department of Physics, Cambridge, MA 02139, USA.*
[2]*Geballe Laboratory for Advanced Materials, Stanford University, Stanford, CA 94305, USA.*
[3]*Department of Applied Physics, Stanford University, Stanford, CA 94305, USA.*
[4]*SIMES, SLAC National Accelerator Laboratory, Menlo Park, CA 94025, USA.*
[5]*Skolkovo Institute of Technology, Skolkovo Innovation Center, 3 Nobel Street, Moscow, 143026, Russia.*
[6]*Institute for Theoretical and Applied Electrodynamics, Russian Academy of Sciences, Moscow, 125412, Russia.*
[7]*Institute for Theoretical Physics, University of Heidelberg, Philosophenweg 12, 69120 Heidelberg, Germany.*

Upon excitation with an intense ultrafast laser pulse, a symmetry-broken ground state can undergo a non-equilibrium phase transition through pathways dissimilar from those in thermal equilibrium. Determining the mechanism underlying these photo-induced phase transitions (PIPTs) has been a long-standing issue in the study of condensed matter systems [1]. To this end, we investigate the light-induced melting of a unidirectional charge density wave (CDW) material, $LaTe_3$. Using a suite of time-resolved probes, we independently track the amplitude and phase dynamics of the CDW. We find that a quick ($\sim 1$ ps) recovery of the CDW amplitude is followed by a slower reestablishment of phase coherence. This longer timescale is dictated by the presence of topological defects: long-range order (LRO) is inhibited and is only restored when the defects annihilate. Our results provide a framework for understanding other PIPTs by identifying the generation of defects as a governing mechanism.

The understanding of equilibrium phase transitions caused by spontaneous symmetry breaking is a hallmark achievement of 20th-century physics. When these transitions are induced by adiabatic cooling from a disordered to an ordered phase, they are marked by a diverging correlation length and correlation time of equilibrium fluctuations at the transition temperature, $T_c$ [2]. Much less is understood about non-adiabatic transitions, or quenches, where fluctuations are not expected to exhibit a diverging correlation length and time, preventing the onset of LRO. This absence of critical behavior is often linked to the creation of topological defects in the ordered phase. The conventional framework for treating non-adiabatic transitions, Kibble-Zurek theory [3, 4], suggests that as the system is quenched through a phase transition from a disordered state, topological defects are generated as a result of the simultaneous emergence of the ordered phase in disconnected regions of space. While such a picture is supported by experiments, for example, in liquid crystals [5, 6] and $^3$He [7, 8], it has so far seen only limited experimental support in a broad new class of non-adiabatic transitions induced in ordered systems by photoexcitation [9, 10].

PIPTs present a unique platform whereby non-adiabatic transitions can be studied. They have emerged as an intense research field in recent decades [1] as a consequence of the technological advances offered by ultrafast lasers. During these transitions, the initial state appearing immediately after photoexcitation, from which order recovers, is far from equilibrium. Moreover, topological defects in this case are not necessarily generated through a complete melting of the broken symmetry phase, but may also arise within the ordered state as a result of spatially localized absorption of high-energy photons.

Materials that exhibit a unidirectional incommensurate CDW are well-suited for investigating PIPTs. Topological defects in these systems, such as dislocations, have been classified theoretically [11, 12] and are thought to play a negligible role in an equilibrium metal-to-CDW transition [13]. Indeed, if a sample is adiabatically cooled below $T_c$, a resolution-limited diffraction peak appears [14–16]. This observation indicates that the phase coherence extends macroscopically without impedance from topological defects, which, when present, reduce the correlation length and disrupt LRO. By contrast, pre-vious studies on PIPTs in unidirectional CDW systems have hinted at the existence of topological defects [9, 10], but more direct probes are needed to elucidate how their presence affects the order parameter dynamics.

* These authors contributed equally.
† Present address: Center for Free-Electron Laser Science, DESY, Notkestraße 85, 22607 Hamburg, Germany.
‡ Present address: University of Louisville, Department of Physics and Astronomy, Louisville, KY 40292, USA.
§ Present address: Geballe Laboratory for Advanced Materials, Stanford University, Stanford, CA 94305, USA.
¶ Present address: Department of Physics, Harvard University, Cambridge, MA 02138, USA.
** gedik@mit.edu

In this work, we use three different time-resolved probes to gain insight into the light-driven phase transition kinetics (see Methods). In each probe, an incident pump pulse perturbs or melts the CDW, and a delayed probe pulse is utilized to measure the ensuing dynamics of the relevant observable. We employ ultrafast electron diffraction (UED) to probe the long-range density correlations [17], while using transient reflectivity and time- and angle-resolved photoemission spectroscopy (tr-ARPES) to track the CDW gap amplitude [9, 18–22]. Transient reflectivity has the advantage that it possesses the highest temporal resolution and signal-to-noise ratio among the probes used, enabling us to additionally investigate the coherent response from collective excitations. The main benefit of tr-ARPES lies in its energy and momentum resolution; hence, it can directly probe the relevant gap dynamics. Each of the three techniques provides a unique perspective, allowing us to gain a comprehensive view of the PIPT.

The material we choose to study is the unidirectional CDW compound $LaTe_3$. It has a simple phase diagram [16], providing a clean platform to explore the effect of photoexcitation. Its layered structure, whose $b$-axis lies out-of-plane [23], makes it susceptible to CDW order that forms below an estimated transition temperature of $\sim 670\,\mathrm{K}$, with an associated gap of $2\Delta \approx 750\,\mathrm{meV}$ [24]. Because of a small in-plane anisotropy in the material, the CDW forms solely along the crystallographic $c$-axis, with an incommensurate wavevector $\mathbf{q}_0 \approx \frac{2}{7}\mathbf{c}^*$, where $\mathbf{c}^*$ is the reciprocal unit vector [25]. The high value of $T_c$ ensures that, in the course of the PIPT, the transient lattice temperature is maintained below $T_c$ despite laser heating (Fig. S2 and Supplementary Note 2).

We first describe the UED experiments, which monitor the structural modulation through the intensity and width of diffraction peaks. These experiments, carried out in a transmission geometry, are sensitive to both the amplitude and phase coherence of the CDW [12, 26]. Figure 1(a) shows an equilibrium electron diffraction cut along the $(3\ 0\ L)$ line at room temperature. Superlattice peaks, characteristic of CDW formation, are indicated by arrows, and signify the presence of LRO.

Following photoexcitation at $t = 0$, the integrated intensity of the superlattice peak initially decreases within $\sim 1\,\mathrm{ps}$ (Fig. 1(b)), a timescale limited by the temporal resolution of our setup [27]. The intensity then recovers to a quasi-equilibrium value. Meanwhile, the full-width at half maximum (FWHM) also broadens by several times its equilibrium value and subsequently narrows (Fig. 1(b) insets and (c)). The peak broadening observed here is a signature of a loss of LRO, which, in turn, requires the appearance of topological defects in high concentrations [17, 28], a non-trivial consequence of the PIPT. In the example presented in Fig. 1(b), the maximum value of the FWHM implies a CDW correlation length of less than $\sim 10$ crystallographic unit cells (Fig. S5(a)). Based on these estimates and the assumption that the defects are two-dimensional, we calculate that for every two photons absorbed, approximately one defect/anti-defect pair is created (see Supplementary Note 6).

We next study how the time evolution of the superlattice peak changes with excitation density, $F$, shown in Fig. 2(a), where $F$ is quoted in terms of absorbed photons per unit volume (see Supplementary Note 4). Beyond a critical value, $F_c \approx 2.0 \times 10^{20}\,\mathrm{cm}^{-3}$, the CDW melts, as the peak becomes indistinguishable from noise after photoexcitation (Figs. 2(a) and S4(a)). We estimate that the critical excitation density, $F_c$, corresponds to a defect every $\sim 6$ crystallographic unit cells, a length-scale below which it is no longer appropriate to define the CDW with the wavevector $\mathbf{q}_0 \approx \frac{2}{7}\mathbf{c}^*$.

To understand how the CDW is reestablished after photoexcitation, we focus on the recovery timescale of the integrated intensity shown in Figs. 2(a) and 4(a). Most significantly, the characteristic time it takes for the peak to recover to quasi-equilibrium increases with excitation density, reaching $5.5\,\mathrm{ps}$ for the largest value of $F$. In general, a transient reduction in the integrated intensity can be caused by four different factors: (i) a suppression of the CDW amplitude; (ii) the excitation of phase modes (phasons) [26, 29]; (iii) a decrease of out-of-plane CDW correlation length [12], which we were not able to access in the transmission geometry of our experiment; and (iv) scattering from defect cores, which redistribute intensity across the entire Brillouin zone [17]. The latter two factors are controlled by the concentration of topological defects. The increased population of phasons originate partially from the temperature rise due to the laser pulse, but can also stem from defect motion [9]. These factors suggest that the dynamics of topological defects may be intimately tied to the recovery of superlattice peaks. On the experimental side, this link is based on the following observation: the time evolution of the integrated intensity closely tracks that of the CDW correlation length (Figs. S5, 1(b), and Supplementary Note 6). Therefore, we infer that it is the reestablishment of CDW phase coherence that dictates the recovery timescale of the superlattice peak.

The CDW amplitude – the first factor in the above list – recovers on a quicker timescale than the phase coherence, as we demonstrate in the following. For this purpose, we examine the structural Bragg peaks. In theory, their dynamics reflect the response of only the CDW amplitude, whereas superlattice peaks additionally retain information about the phase coherence [26]. In Fig. 1(b), we show that when the CDW is suppressed, the Bragg peak first intensifies; subsequently, it weakens due to the Debye-Waller factor. This initial intensification signifies a reduction of the CDW amplitude, as distorted atoms return to their high symmetry positions. Significantly, the Bragg peak enhancement disappears on a quicker timescale than it takes for the superlattice peak to regain its intensity (Fig. S6 and Supplementary Note 7), suggesting that the CDW amplitude and phase coherence recover at different rates.

To confirm this picture, we turn to transient reflectivity and tr-ARPES measurements, which can probe the fast evolution of CDW amplitude [9, 18–22] with a finer temporal resolution (see Methods). In Fig. 2(b), we present the transient reflectivity results where two components in each trace are visible: an incoherent response arising from the excitation and relaxation of quasiparticles, and an oscillating coherent response predominantly from the 2.2 THz CDW amplitude mode (AM) [20]. From the coherent AM response, we determined that the CDW melts at the critical excitation density, $F_c$, consistent with the UED measurement (Fig. S4 and Supplementary Note 5). However, the CDW recovery timescale, extracted from the incoherent response, is much quicker when compared to that of the UED superlattice peak (Figs. 2(a)–(b) and 4(a)). As the incoherent response is known to be a sensitive probe of the CDW gap size [9, 18–20], we again infer that the amplitude recovers on a faster timescale than the phase coherence.

To further investigate the amplitude restoration in a momentum- and energy-resolved fashion, we use tr-ARPES to probe the gap dynamics. Figure 3(a) shows a segment of the Fermi surface excited with a density above $F_c$. Following photoexcitation, spectral weight fills in the gapped portions of the Fermi surface. In particular, Fig. 3(b) shows the time evolution of the momentum-integrated intensity of a representative gapped region (orange box in Fig. 3(a)), where the Fermi level, $E_F$, lies approximately at the center of the CDW gap [30]. The gap size decreases and is subsequently restored to its quasi-equilibrium value within a couple of picoseconds (Fig. S7 and Supplementary Note 8). To characterize the gap recovery more quantitatively, we plot in Fig. 3(c) the time evolution of in-gap spectral weight obtained by integrating the intensity over an energy window of $\pm 0.1$ eV around $E_F$ (orange box in Fig. 3(b)). After fitting the recovery with an exponential decay, the time constant obtained is less than 1.1 ps for all excitation densities measured, consistent with the transient reflectivity results (Fig. 4(a)). We thus identify $\sim$ 1 ps as the characteristic timescale for the restoration of the CDW amplitude, whereas the phase coherence takes up to 5.5 ps to recover at the highest excitation density measured (Fig. 4(a)).

Taken together, our investigations of $LaTe_3$ are consistent with the interpretation sketched in Fig. 4(b)–(e). In this picture, the laser pulse excites energetic quasiparticles, which through recombination, create topological defects. In the illustration, these defects are depicted as CDW dislocations [11, 28, 29]. In the meantime, the CDW gap and hence the amplitude of the CDW order parameter decreases, while the long-range phase coherence is suppressed or destroyed (Fig. 4(c)). Within $\sim$ 1 ps, the CDW amplitude recovers to quasi-equilibrium. Meanwhile, the long-range phase coherence is not fully restored (Fig. 4(d)). It takes several more picoseconds or longer, depending on the excitation density, for the defects to annihilate and for the CDW phase coherence to set in (Fig. 4(e)).

An important implication of Fig. 4(a) is that the phase coherence takes longer to recover with a higher concentration of topological defects (orange line). While pinpointing the exact reason behind this relationship requires a detailed theoretical treatment, here we limit ourselves to mentioning two possible mechanisms. If the recovery is determined by the annihilation of pairs of two-dimensional topological defects, then the presence of a large number of defects disturbs the coupling between CDWs in adjacent planes. Hence, this renormalized out-of-plane coupling reduces the restoring force that brings together defect/anti-defect pairs. A related possibility is that larger concentrations of photo-induced defects renormalize down the effective $T_c$ [2] to a value close to that of the laser-heated sample in UED (see Supplementary Note 2), and as a result, the restoration of the CDW order exhibits critical slowing down associated with the proximity to the renormalized $T_c$ [2].

The synergy of the three time-resolved probes used in this work has provided a uniquely comprehensive view of the photo-induced phase transition in a symmetry-broken state. Across the multiple techniques, a consistent picture is obtained where a quick recovery of the CDW amplitude is followed by a slower restoration of phase coherence. Topological defects are preeminent in this regard, inhibiting the reestablishment of LRO in the non-equilibrium setting. A transition driven by photoexcitation, where topological defects are generated with photons in the ordered phase, therefore represents a distinctive framework under which non-adiabatic transitions can be instigated. These results pave the way to future studies on non-equilibrium defect-mediated transitions and the optical manipulation of topological defects in other ordered states of matter.

## METHODS

**Sample preparation**. Single crystals of $LaTe_3$ were grown by slow cooling of a binary melt [23]. For the UED measurement, $LaTe_3$ was mechanically exfoliated to thickness between 10 and 30 nm and then transferred to a 10 nm thick silicon nitride TEM window in an inert gas environment (see Fig. S1(a) and Supplementary Note 1). For tr-ARPES and transient reflectivity measurements, bulk single crystals were cleaved in ultrahigh vacuum ($< 1 \times 10^{-10}$ torr) and in inert gas, respectively, to expose a pristine surface.

**Ultrafast electron diffraction**. The 1038 nm (1.19 eV) output of a commercial Yb:KGW regenerative amplifier laser system (PHAROS SP-10-600-PP, Light Conversion) operating at 250 kHz was split into pump and probe branches. The pump branch was focused onto the sample at room temperature, while the probe branch was frequency quadrupled to 260 nm (4.78 eV) and focused onto a gold-coated sapphire in high vacuum ($< 4 \times 10^{-9}$ torr) to generate photoelectrons. These electrons were accelerated to 26 kV in a dc field and focused with a solenoid before diffracting from $LaTe_3$ in a transmission geometry, making visible the $K = 0$ diffraction plane. Diffracted electrons were incident on an aluminum-coated phosphor screen (P-46), whose luminescence was recorded by a commercial intensified charge-coupled device (iCCD PI-MAX II) operating in shutter mode. A pulse picker was used to tune the laser repetition rate from 0.5 to 250 kHz used in the measurement. The operating temporal resolution was 1 ps with 1000 to 4000 electrons per pulse.

**Transient reflectivity**. The 780 nm (1.59 eV) output of a commercial Ti:sapphire regenerative amplifier laser (Wyvern 500/1000, KMLabs) operating at 30 kHz was split into pump and probe branches. The probe branch was focused onto a sapphire crystal to generate a white light continuum (500 to 700 nm). Both pump and probe pulses were focused onto the sample surface, which was held at room temperature, at near-normal incidence with parallel polarization. The pump wavelength was 780 nm (1.59 eV), while the probe wavelength was 690 nm (1.80 eV). The reflected probe beam was directed to a monochromator and photodiode for lock-in detection. The overall temporal resolution, as determined from the pump-probe cross-correlation, was 70 fs.

**Time-resolved ARPES**. The same laser system employed in UED measurements was used for tr-ARPES. The pump branch was passed to a commercial optical parametric amplifier (ORPHEUS, Light Conversion) to generate a 720 nm (1.72 eV) output used to photoexcite the sample. The probe branch was first frequency-tripled to 346 nm (3.58 eV) and then focused into a hollow fiber filled with xenon gas (XUUS, KMLabs) to generate the 9th harmonic at 115 nm (10.75 eV). The resulting XUV pulse was passed through a custom-built grating monochromator (McPherson OP-XCT) to minimize pulse width broadening and to enhance throughput efficiency [31], before it was focused onto the sample, which was held at 15 K for optimal resolution (see Supplementary Note 9). Photoelectrons were collected by a time-of-flight detector (Scienta ARTOF 10k), which made simultaneous measurements of the energy and in-plane momenta possible. The operating temporal resolution was 230 fs at a laser repetition rate of 250 kHz. The energy resolution for the measurements was 50 meV.

## ACKNOWLEDGEMENTS

We acknowledge helpful discussions with S. Brazovskii, P.A. Lee, J. Ruhman, B. Skinner, A. Krikun, W.H. Zurek, and D. Chowdhury. We acknowledge support from the U.S. Department of Energy, BES DMSE (experimental setup and data acquisition), from the Gordon and Betty Moore Foundation's EPiQS Initiative grant GBMF4540 (data analysis and manuscript writing), Army Research Office (equipment support for the tr-ARPES), and the Skoltech NGP Program (Skoltech-MIT joint project) (theory). Y.Q.B. and P.J.-H. acknowledge support by the Center for Excitonics, an Energy Frontier Research Center funded by the U.S. Department of Energy, Office of Science, Office of Basic Energy Sciences, under award No. DESC0001088, as well as the Gordon and Betty Moore Foundation's EPiQS Initiative through grant GBMF4541 (sample preparation and characterization). Work at Stanford was supported by the U.S. Department of Energy, Office of Basic Energy Sciences, under Contract No. DE-AC02-76SF00515 (sample growth and characterization). P.W. was supported in part by the Gordon and Betty Moore Foundation's EPiQS Initiative through grant GBMF4414. E.B. acknowledges support by the Swiss National Science Foundation under fellowship P2ELP2-172290.

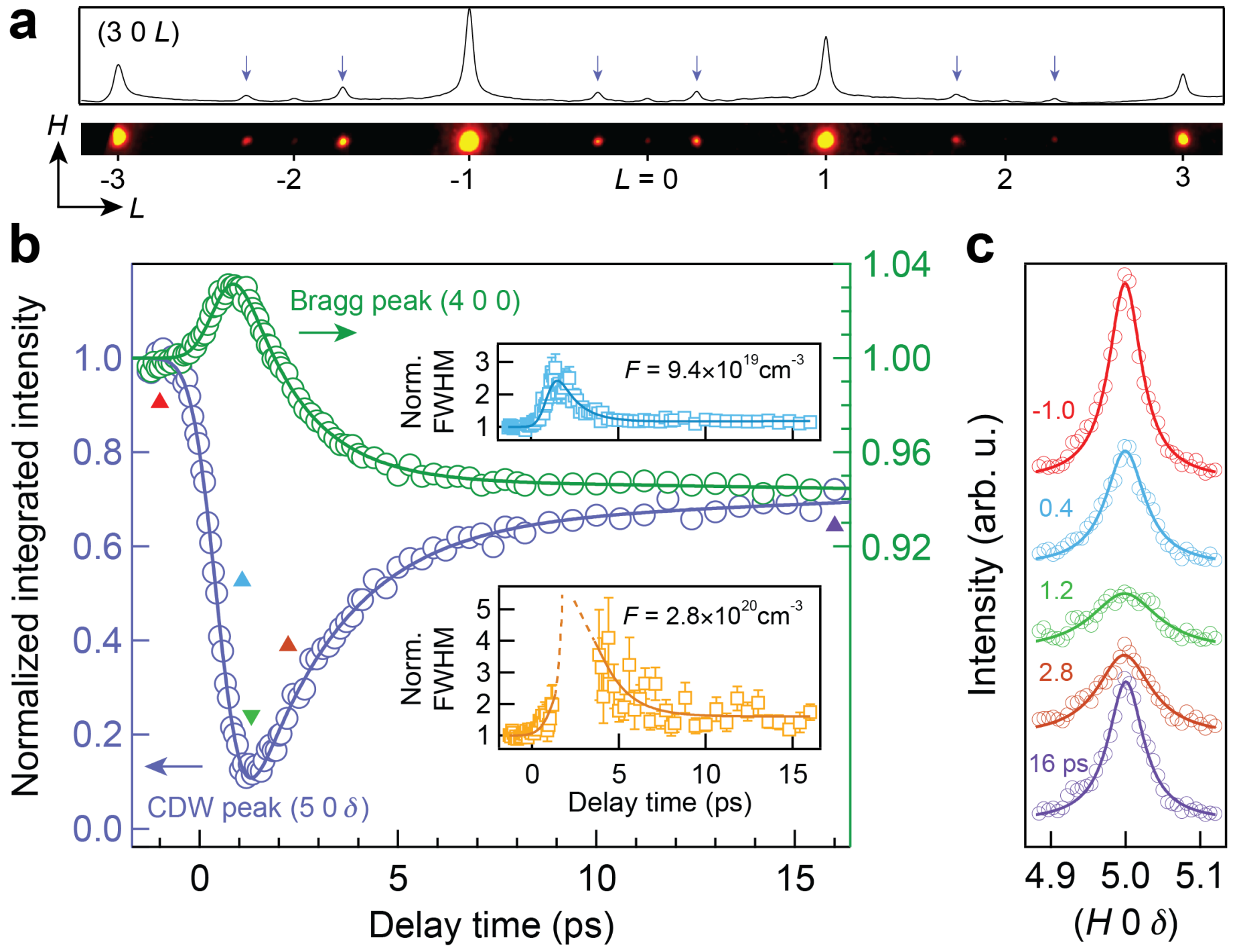

FIG. 1. **Time evolution of electron diffraction after photoexcitation.** (a) Room temperature static electron diffraction cut along the (3 0 $L$) line showing both structural Bragg peaks and CDW superlattice peaks. Superlattice peaks are indicated by arrows. The line cut is obtained by integrating the colored strip shown below along the $H$ direction. A full diffraction pattern is shown in Fig. S1(b). (b) Time evolution of integrated intensities of the (4 0 0) Bragg peak and the (5 0 $\delta$) superlattice peak after photoexcitation with a femtosecond light pulse at an excitation density of $9.4 \times 10^{19}\,\mathrm{cm}^{-3}$ (see Supplementary Note 4). Intensities are normalized to values before the arrival of the light pulse. Statistical error bars are smaller than the marker size. (Insets) The time evolution of the full-width at half maximum (FWHM) of the superlattice peak, normalized to values before photoexcitation, showing significant broadening through the CDW transition. Error bars are fitting uncertainties. All solid curves in (b) are fits to a phenomenological relaxation model (Supplementary Notes 3 and 7), while dashed lines in the lower inset are extrapolated to regions where the peak vanishes. (c) Snapshots of the superlattice peak at selected time delays, indicated by the triangles in (b). The transient broadening is isotropic along both $H$ and $L$ directions, and the line profiles shown are along $H$, from which FWHMs are computed by fitting to a Lorentzian function (solid curves).

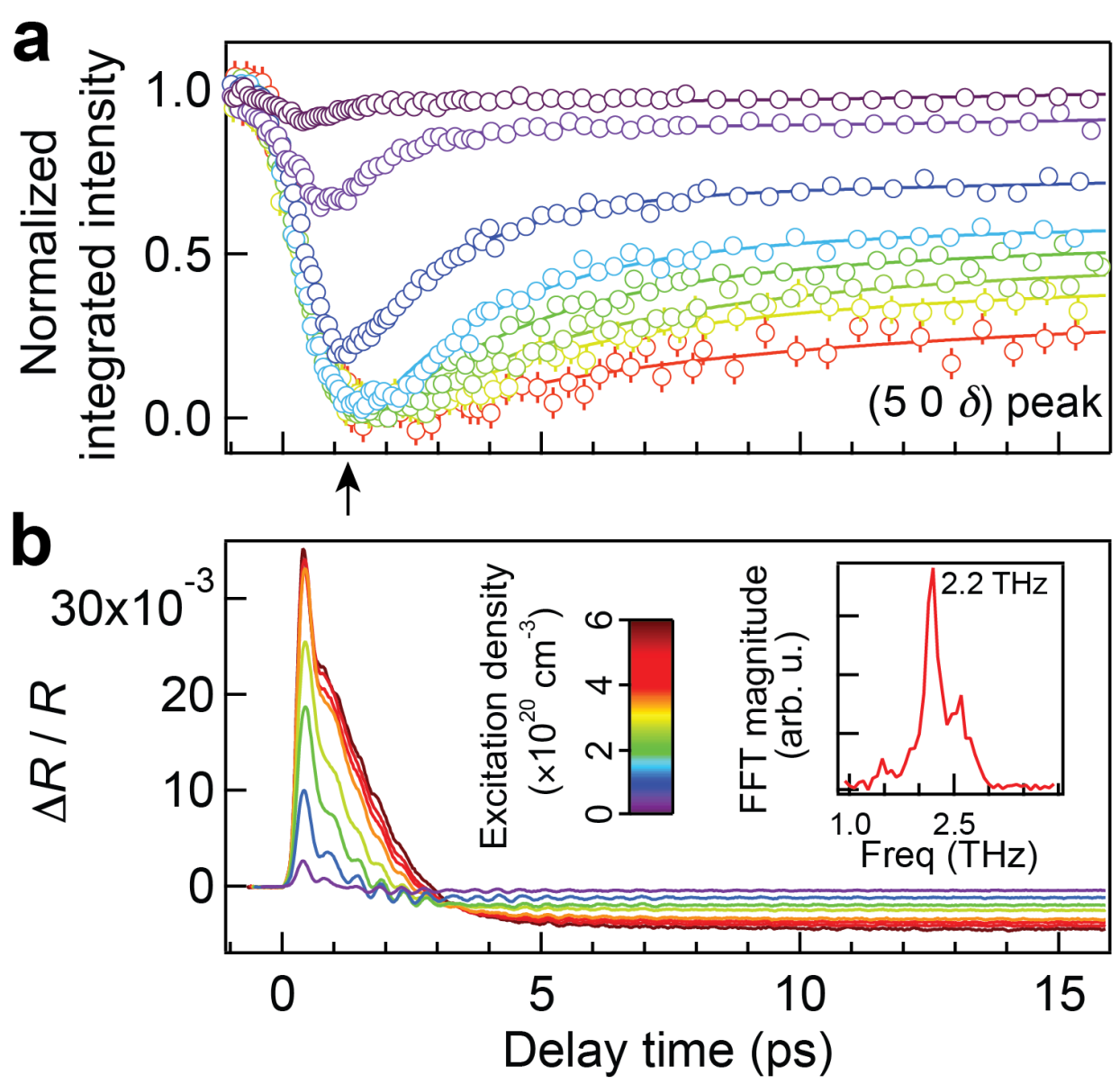

FIG. 2. **Dependence of CDW diffraction intensity and optical reflectivity on excitation density.** (a) Time evolution of integrated intensity of the (5 0 $\delta$) superlattice peak upon photoexcitation at different excitation densities. The color scale is the same as used in (b). Solid curves are fits to a phenomenological relaxation model (Supplementary Note 3). The arrow indicates the time delay at which the intensity is plotted against the excitation density in Fig. S4(a). (b) Transient reflectivity as a function of delay time at different excitation densities. (Inset) Fourier transform of the oscillatory component measured at an excitation density of $4.1 \times 10^{19}\,\mathrm{cm}^{-3}$ (see Supplementary Note 5).

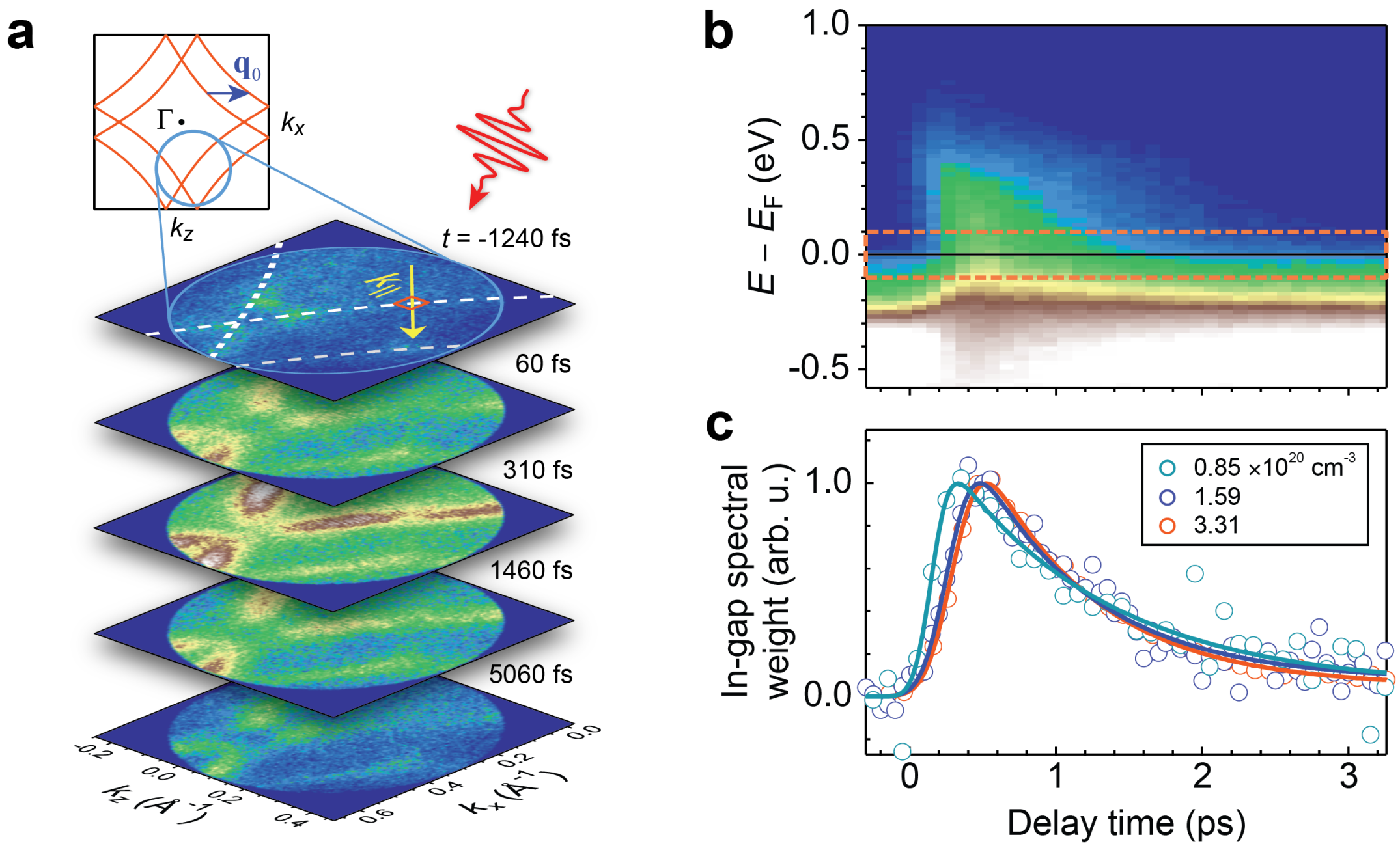

FIG. 3. **tr-ARPES spectra showing CDW gap dynamics.** (a) (Top) Tight-binding plot of the normal-state Fermi surface formed by the Te $p$-bands in the first Brillouin zone. A circle shows the probed part of the Fermi surface. An arrow marks the CDW wavevector $\mathbf{q}_0$. (Bottom) A section of the Fermi surface through the photoexcitation process with an excitation density of $3.31 \times 10^{20}\,\mathrm{cm}^{-3}$. Intensities are integrated over $\pm 10\,\mathrm{meV}$ around $E_F$. Cuts along the $k_{||}$ line (yellow arrow) are shown in Fig. S7(a)–(e). (b) The time evolution of the gapped region highlighted with the orange box drawn in the $t = -1240\,\mathrm{fs}$ slice of (a). (c) The tr-ARPES time evolution of the in-gap spectral weight obtained by integrating the intensity over $\pm 0.1\,\mathrm{eV}$, as outlined by the orange box in (b). The three traces correspond to different excitation densities, all normalized between 0 and 1. Curves are fits to a single-exponential relaxation model (Supplementary Note 3).

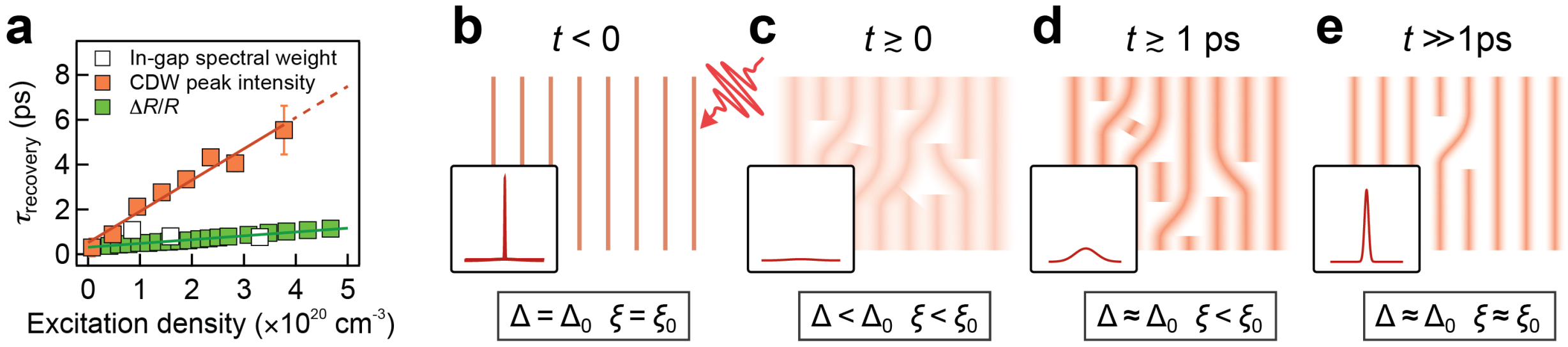

FIG. 4. **Summary of CDW recovery timescales and dynamics.** (a) Characteristic recovery times across three probes as a function of excitation density. The orange squares are taken from UED, the green from transient reflectivity and the white from tr-ARPES. Error bars, when larger than the symbol size, denote fitting uncertainties. Lines are linear fits to the data where the dashed segment denotes a linear extrapolation. (b)–(e) Schematic illustration of the CDW evolution after photoexcitation. In each image, the unidirectional charge density modulation is depicted as stripes in real space. Stripe brightness indicates the strength of the CDW amplitude and smearing represents phase excitations. A cartoon of the CDW diffraction peak is presented in the lower left corner. $\Delta$ and $\xi$ denote CDW amplitude and correlation length, respectively; $\Delta_0$ and $\xi_0$ are values at equilibrium. (b) Before photoexcitation, the CDW amplitude is large and the CDW is long-range ordered. The corresponding superlattice diffraction is represented by a narrow-width peak. (c) Following photoexcitation, the CDW amplitude is suppressed and topological defects are formed. These effects lead to a reduction in the integrated intensity and a broadening of the peak width. (d) After $\sim$1 ps, the CDW amplitude is largely restored, while defects persist. The diffraction peak remains broad due to the presence of these topological defects. (e) Many defects annihilate at a further time delay though a non-zero defect concentration remains. The superlattice peak significantly narrows as the phase coherence sets in.

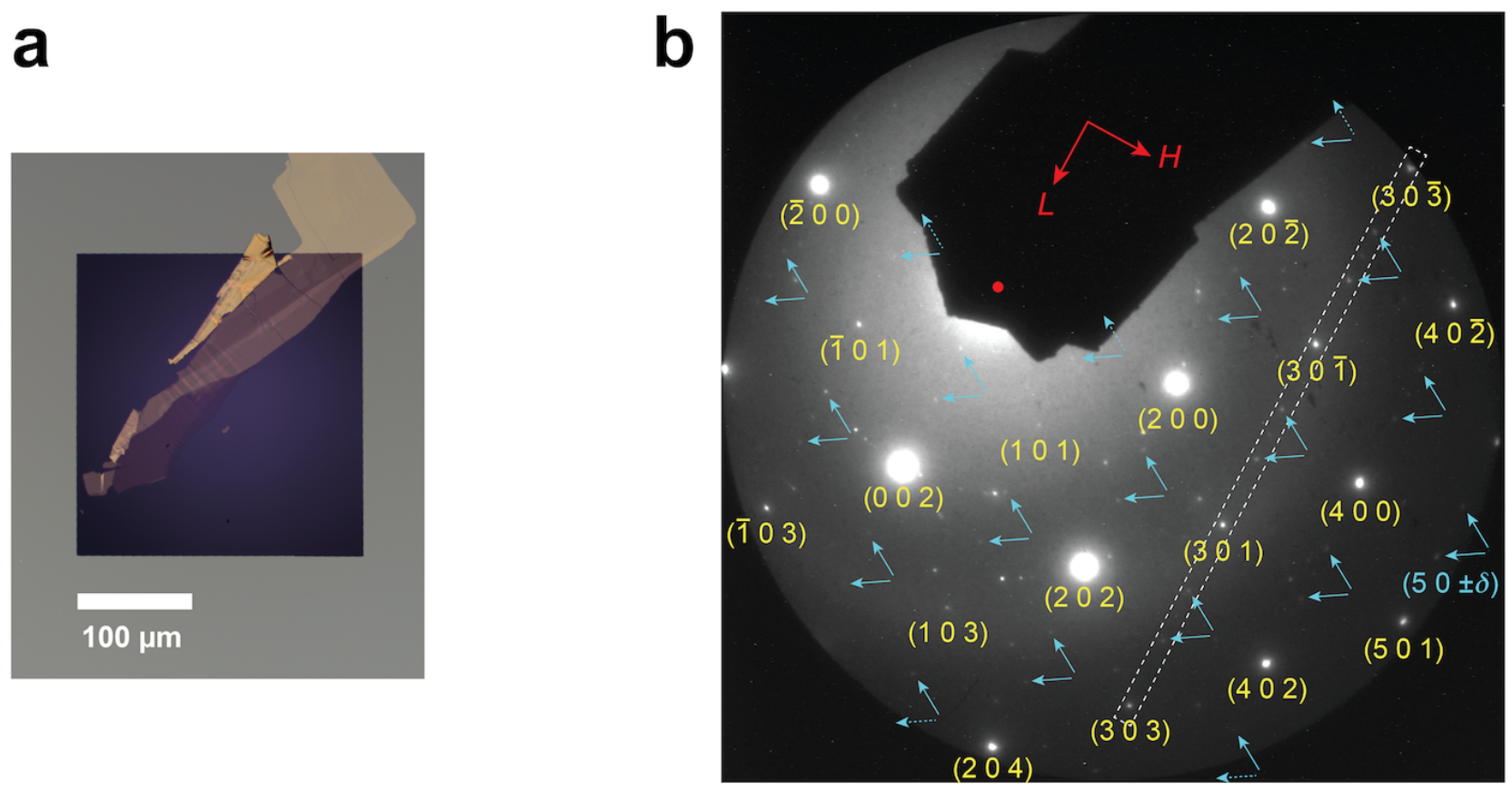

FIG. S1. (a) Optical image of the UED sample on an amorphous silicon nitride membrane. (b) Full static electron diffraction pattern of $LaTe_3$, where the red dot marks the position of the undiffracted incident beam. Blue arrows mark the superlattice peaks. The boxed region is shown in Fig. 1(a).

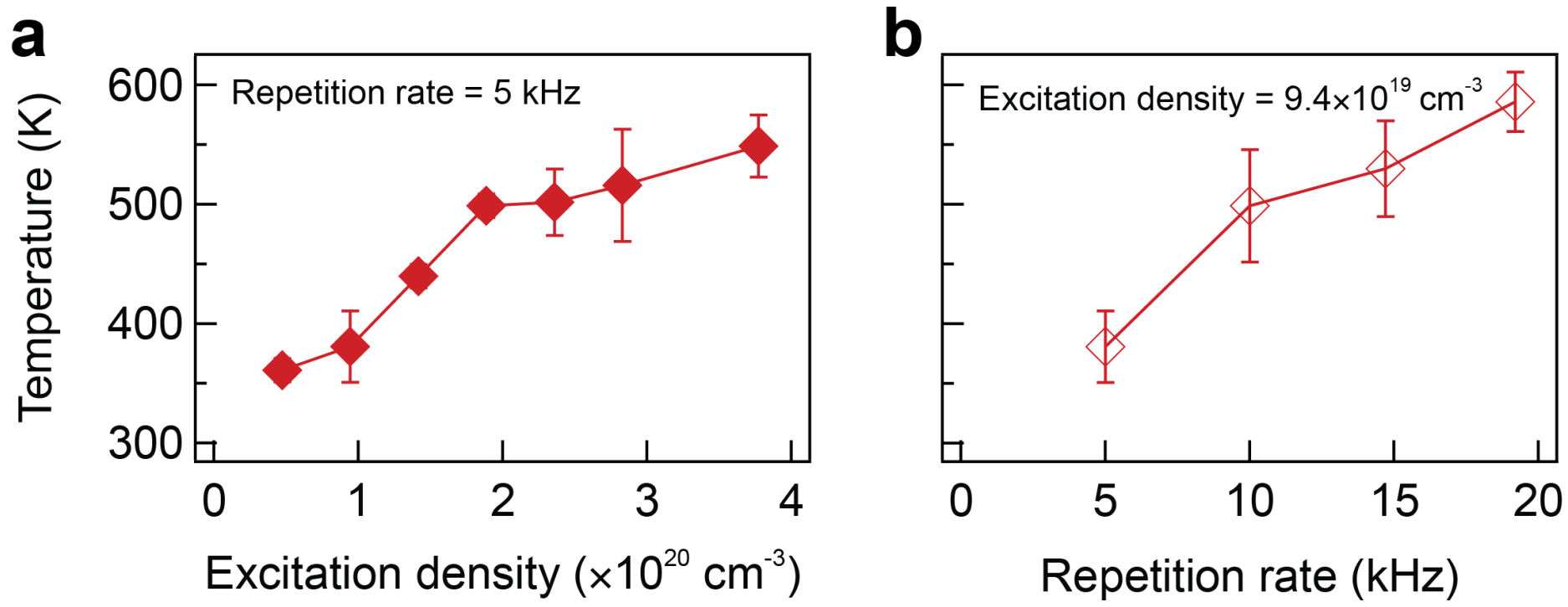

FIG. S2. (a) Estimated steady-state temperature of the sample as a function of excitation density at a fixed 5 kHz laser repetition rate. (b) Estimated steady-state temperature of the sample as a function of repetition rate at a fixed excitation density of $9.4\times10^{19}\,\mathrm{cm}^{-3}$.

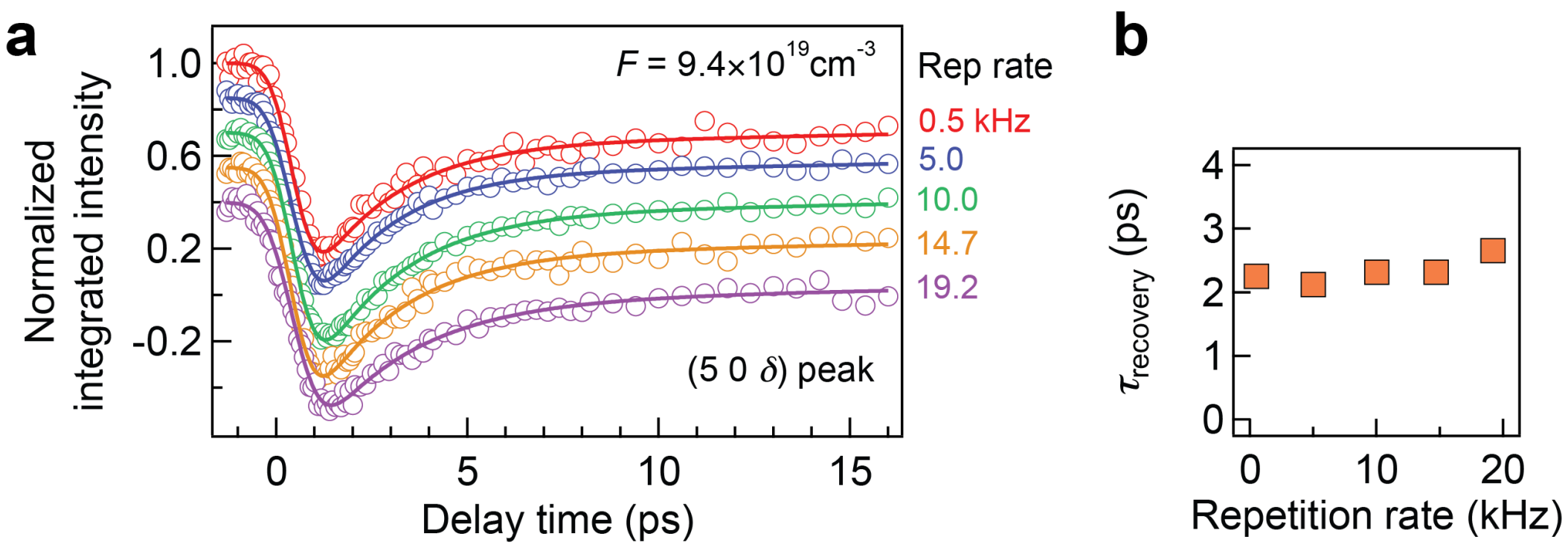

FIG. S3. (a) Time evolution of the integrated intensity of the (5 0 $\delta$) superlattice peak at several laser repetition rates with a fixed excitation density of $9.4 \times 10^{19}\,\mathrm{cm}^{-3}$. Each trace has been vertically offset by $-0.15$ for clarity. Solid curves are fits to a single-exponential relaxation model (see Supplementary Note 3). (b) Recovery time constants extracted from the fits in (a).

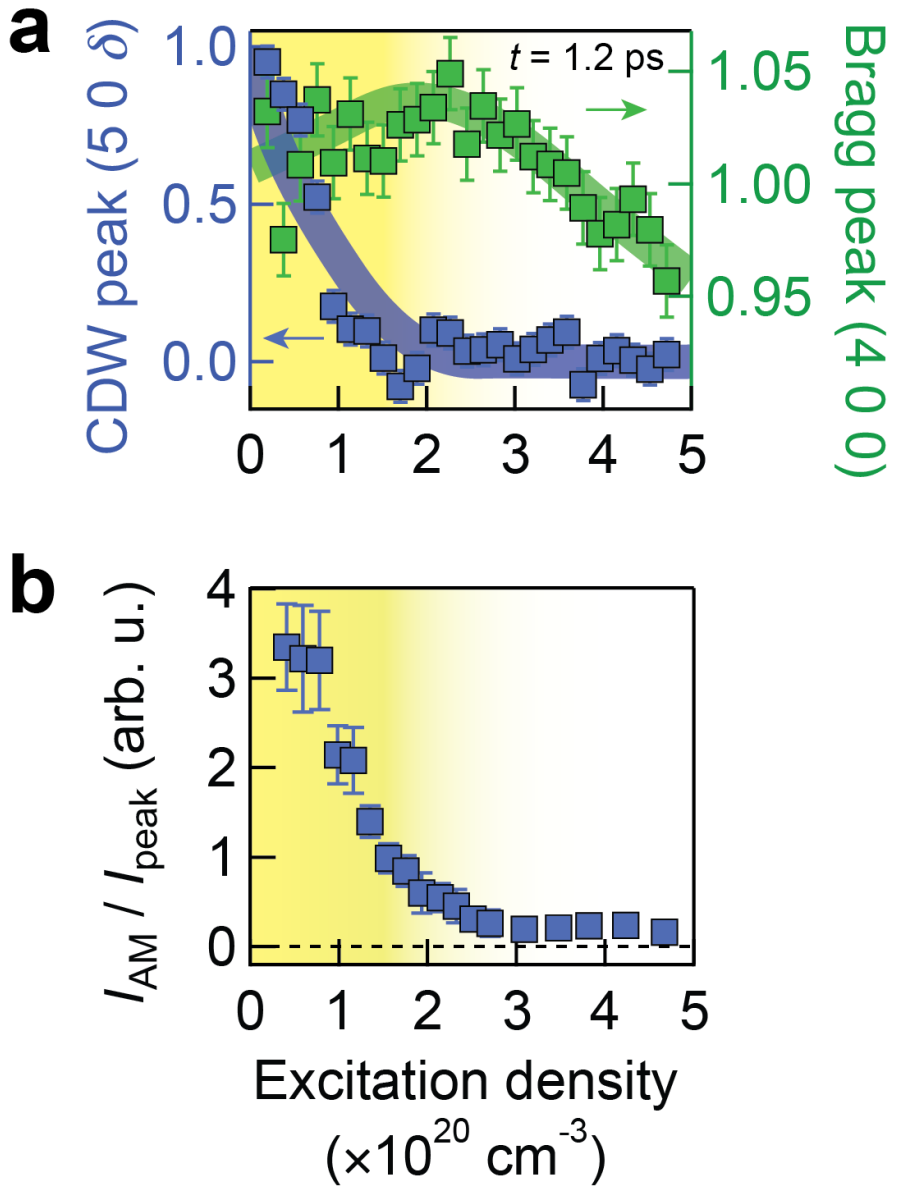

FIG. S4. (a) The normalized (4 0 0) Bragg and (5 0 $\delta$) superlattice peak integrated intensities at a time delay of $t = 1.2\,\mathrm{ps}$ (arrow in Fig. 2(a)) as a function of excitation density. Curves are guides to eye. (b) Normalized amplitude mode (AM) intensity as a function of excitation density. The AM intensity, $I_{\mathrm{AM}}$, has been normalized against the maximum of the incoherent part, $I_{\mathrm{peak}}$, in the transient reflectivity trace (see Supplementary Note 3). In both panels, yellow shaded regions denote the low excitation regime where the CDW has yet to melt. Error bars are derived from fitting uncertainties.

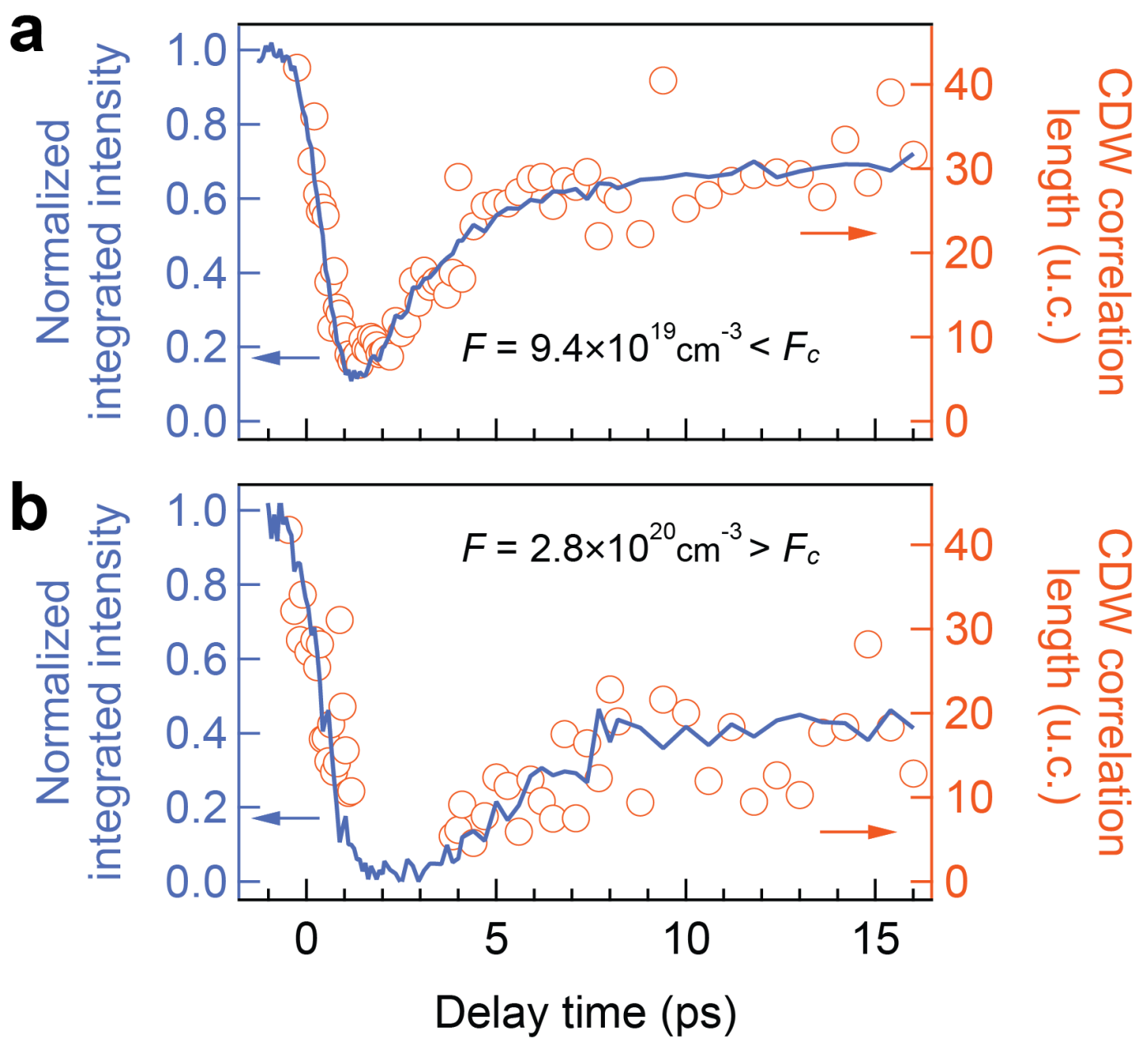

FIG. S5. Time evolution of the integrated intensity of the (5 0 $\delta$) CDW peak (blue curves), overlaid onto the evolution of CDW correlation length (orange circles) obtained from the FWHM shown in the insets of Fig. 1(b). (a) and (b) correspond to two excitation densities below and above the melting threshold, $F_c \approx 2.0 \times 10^{20}\,\mathrm{cm}^{-3}$. Missing circles in (b) from 1 to 4 ps correspond to the time range where CDW peaks are indistinguishable from the background and the FWHM cannot be reliably extracted from fittings. u.c., crystallographic unit cells.

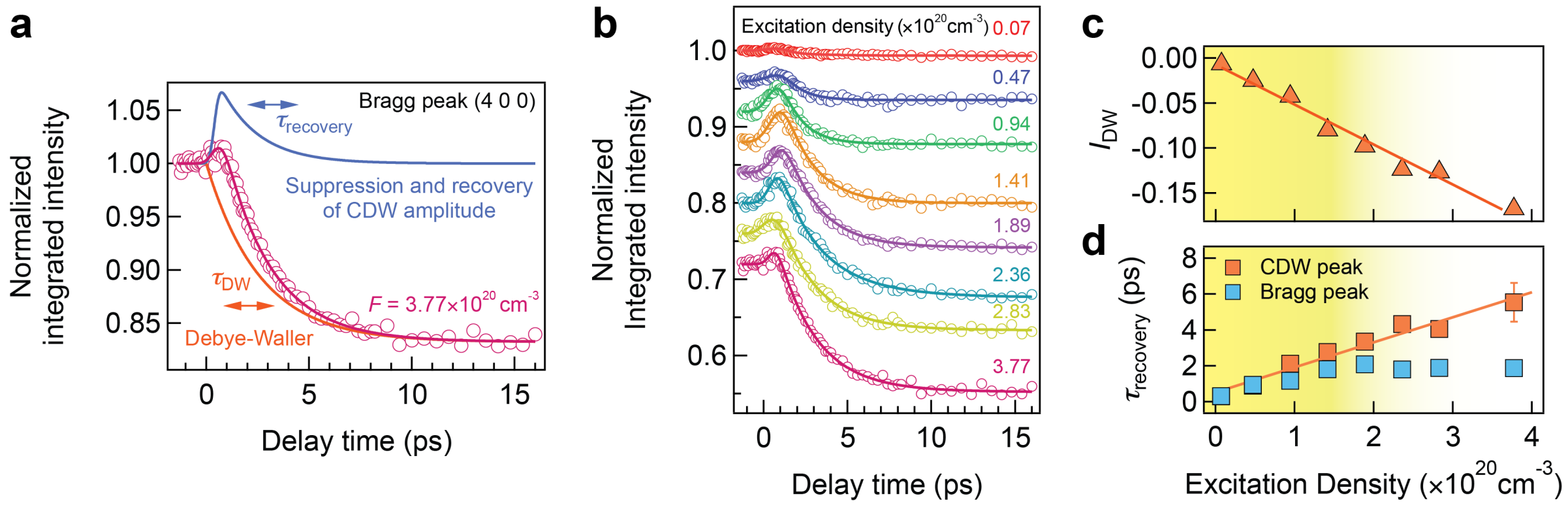

FIG. S6. (a) Time evolution of the integrated intensity of the (4 0 0) Bragg peak at an excitation density of $3.77 \times 10^{20}\,\mathrm{cm}^{-3}$. Blue and orange curves represent two fitted components that account for the change in the CDW amplitude and the Debye-Waller factor, respectively. (b) Time evolution at different excitation densities, with fitted curves superimposed. Each trace is vertically offset by $-0.04$ for clarity. (c) Loss of Bragg peak intensity due to the Debye-Waller factor, $I_{\mathrm{DW}}$, as a function of excitation density (see Eq. (S6)). The solid line is a linear fit. (d) Recovery timescales extracted from Bragg peaks (blue squares), compared with timescales from superlattice peaks (orange squares) reproduced from Fig. 4(a). The yellow-shaded regions in (c) and (d) are the same as those in Fig. S4(a) and (b), where the CDW is only partially suppressed.

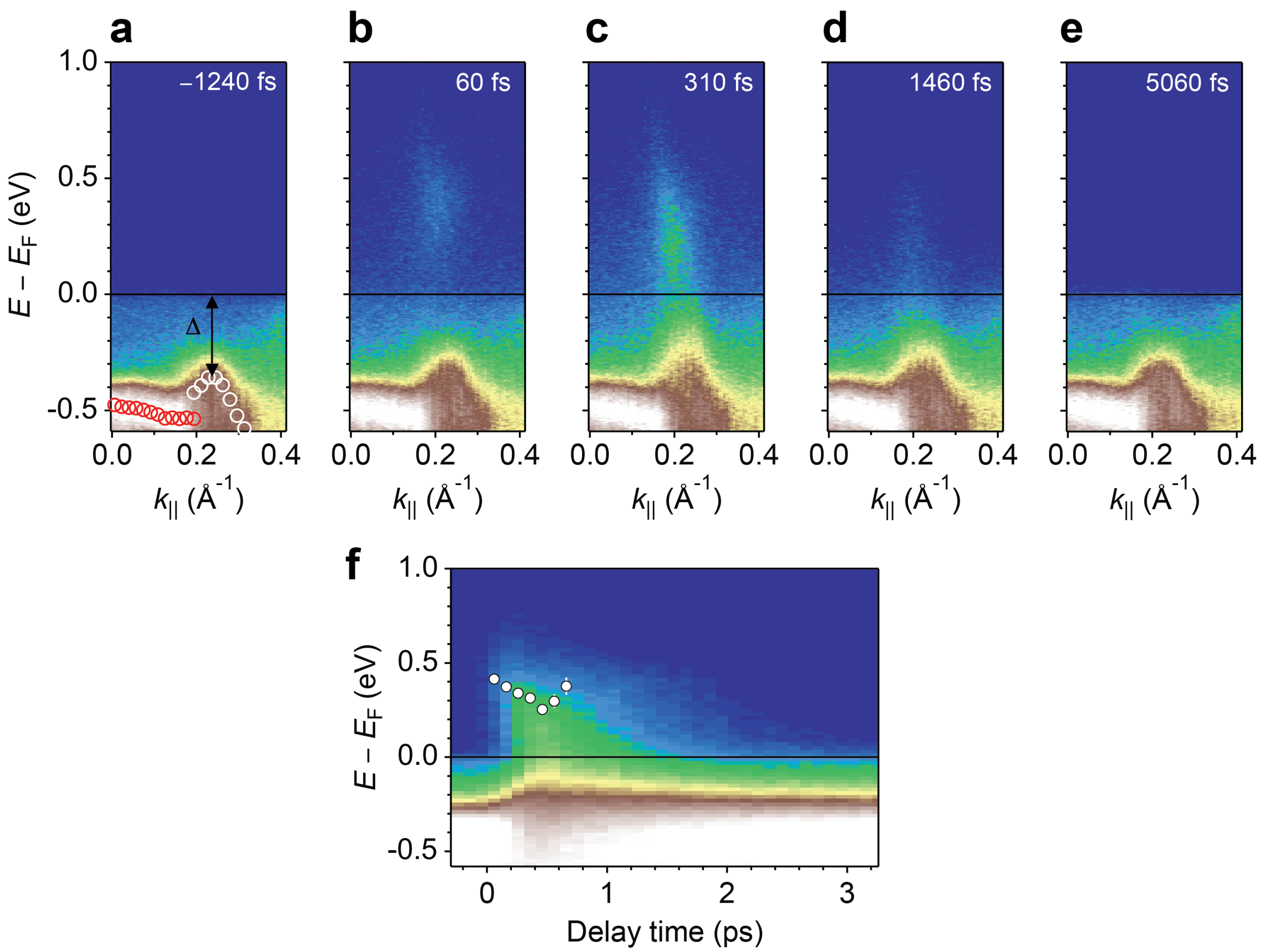

FIG. S7. (a)–(e) Time evolution of the band structure near the Fermi surface after photoexcitation. The momentum cut, $k_{||}$, is indicated in Fig. 3(a). In (a), circles mark the fitted band dispersion of Te $5p_z$ (white) and $5p_y$ (red) bands, where an energy gap $\Delta = 0.36 \pm 0.02\,\mathrm{eV}$ opens at the $5p_z$ band and causes it to back-bend. (f) Reproduced from Fig. 3(b), but with fitted positions of the upper CDW band edge marked by the white dots (see Supplementary Note 9). It uses a different color scale from that in (a)–(e) for better visibility of the band above $E_F$.

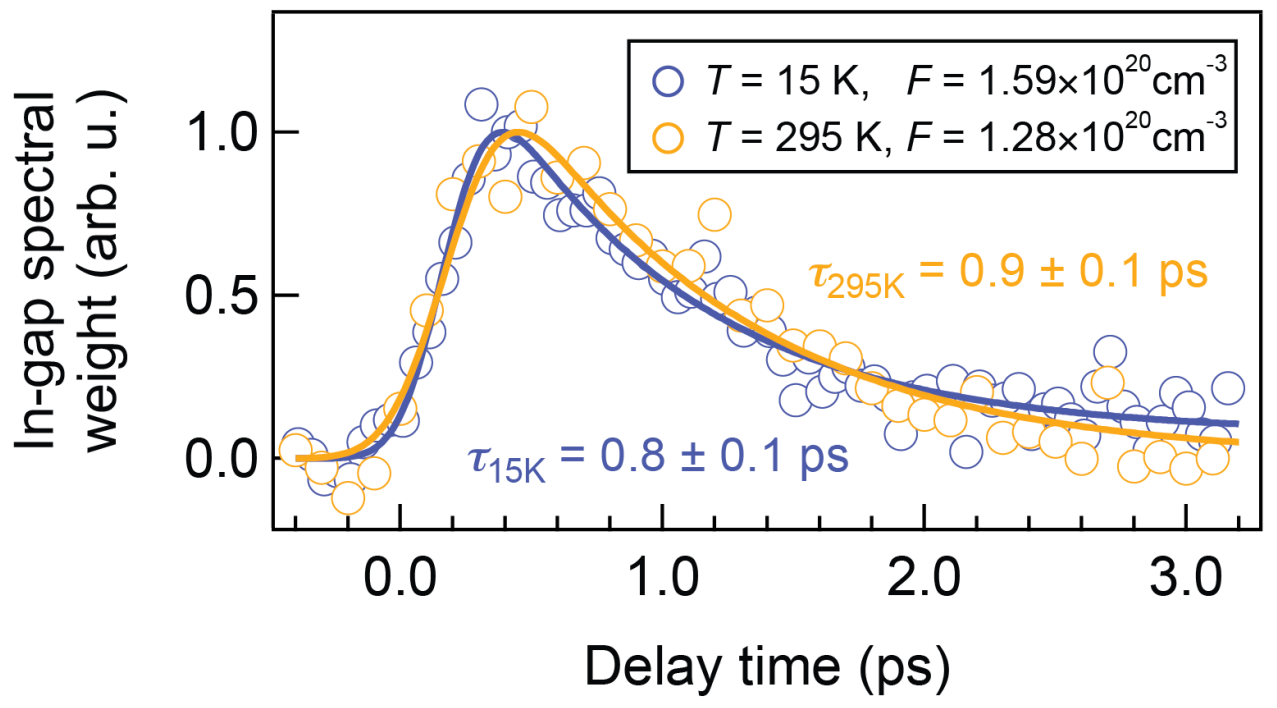

FIG. S8. Time evolution of normalized in-gap spectral weight taken at two different base temperatures and similar excitation densities. The spectral weight is obtained by integrating the intensity over $\pm 0.1\,\mathrm{eV}$ around $E_F$, the orange box in Fig. 3(b).